\documentclass[]{rsos}

\usepackage{amsmath}
\usepackage{lmodern}

\begin{document}

\title{Dynamical Systems in Cosmology: Reviewing An Alternative Approach}

\author{
Nandan Roy$^{1}$, L. Arturo Ure\~na-L\'opez$^{2}$}

\address{$^{1}$Centre for Theoretical Physics \& Natural Philosophy, Mahidol University, Nakhonsawan Campus, Phayuha Khiri, Nakhonsawan 60130, Thailand\\
$^{2}$Departamento de F\'isica, DCI, Campus León, Universidad de Guanajuato, 37150, León, Guanajuato, México}

\subject{xxxxx, xxxxx, xxxx}


\corres{Nandan Roy \\
\email{nandan.roy@mahidol.ac.th}}

\begin{abstract}

Dark energy is one of the deepest puzzles in modern cosmology, and mounting evidence suggests that it is not just a cosmological constant but a genuinely dynamical component. Although cosmology and dynamical systems theory emerged from different disciplines, dynamical systems methods have become essential tools to uncover the qualitative evolution of the universe. The equations governing homogeneous and isotropic cosmologies can be naturally written as systems of ordinary differential equations, making them an ideal arena for dynamical system analysis. This review begins with a sharp, streamlined introduction to the standard dynamical systems toolkit widely used in cosmology. We then move on to alternative formulations based on polar and hyperbolic variable transformations. These approaches unlock powerful new ways to probe a broad spectrum of scalar field dark energy models, to set and constrain initial conditions, and to analyze tracking behavior across wide classes of potentials. The review is self-contained, but consistently directs the reader to more specialized and in-depth treatments where needed.
\end{abstract}

\maketitle

\section{Introduction}

The observational evidence for the late--time accelerated expansion of the Universe stands as one of the most significant discoveries in modern cosmology. Measurements from type Ia supernovae\cite{SupernovaCosmologyProject:1998vns,SupernovaSearchTeam:1998fmf}, the cosmic microwave background \cite{Planck2020}, and large-scale structure surveys \cite{1998ApJ496605E} consistently indicate that nearly seventy percent of the total energy density of the Universe is attributed to an exotic component with negative pressure, commonly referred to as dark energy. Despite its observational success, the physical origin of dark energy remains unknown, motivating the exploration of dynamical alternatives to the cosmological constant.

Among the various proposed models, the scalar field dark energy scenarios ~\cite{amendola2010dark, Bamba:2012cp,copeland2006dynamics, Peebles2003, Armendariz2001, roy2022quintessence, Banerjee:2020xcn, Lee:2022cyh, Krishnan:2020vaf} have received considerable attention due to their theoretical motivation and phenomenological richness. These models naturally arise in effective field theories, string-inspired constructions, and modified gravity frameworks. Depending on the choice of the scalar field potential, they can exhibit a wide range of behaviors, including scaling solutions, tracker dynamics, and late--time attractors capable of driving cosmic acceleration\cite{Bahamonde:2017ize}. However, the highly nonlinear nature of the cosmological field equations typically precludes the existence of exact analytical solutions, necessitating qualitative approaches to study their global dynamics.

The theory of dynamical systems provides a powerful and systematic framework for analyzing cosmological models in this context. By recasting the Einstein field equations and matter conservation laws as autonomous systems of ordinary differential equations, one can investigate the structure of the associated phase space, identify equilibrium points, assess their stability, and determine the generic asymptotic behavior of cosmological solutions. Over the past decades, dynamical systems techniques have been extensively applied in both general relativity and cosmology. A wide range of important models have been analyzed using this approach, including modified gravity theories, scalar–tensor models, Bianchi cosmologies, and scenarios involving non-minimally coupled scalar fields~\cite{coley2003dynamical,Coley:1999uh,copeland2006dynamics,Bahamonde:2017ize,Roy:2015jkm}.

The application of dynamical system techniques to cosmology dates back to the early 1970s, with several pioneering studies~\cite{Collins1971qualitative,bogoyavlenski,shikin1975anisotropic}. These early investigations primarily focused on anisotropic cosmological models rather than the homogeneous and isotropic Friedmann--Lema\^itre--Robertson--Walker framework. Their main objective was to understand the qualitative dynamics and asymptotic behavior of anisotropic universes through phase-space methods. Over subsequent decades, dynamical systems techniques became an important tool for investigating dark energy models and modified theories of gravity. One of the first influential works employing dynamical systems methods to study the accelerated expansion of the Universe was by Copeland, Liddle, and Wands~\cite{Copeland:1997et}, where the formalism was applied to a quintessence scalar field with an exponential potential. Since then, these methods have been extensively used to analyze both canonical and noncanonical scalar field cosmologies. For canonical quintessence fields, a wide variety of scenarios have been studied, including models with minimal and nonminimal couplings to gravity, as well as different choices of scalar-field potentials~\cite{Copeland:1997et,Roy:2014yta,Roy:2013wqa,Urena-Lopez:2011gxx,Capozziello:2003tk,Hrycyna:2007gd,Urena-Lopez:2020npg,Sahoo:2024dgb}. These analyzes revealed the existence of fluid-dominated, scaling, and tracker solutions capable of describing different phases of cosmic evolution. Corresponding studies for non-canonical phantom fields have also uncovered a rich structure of critical points and attractor solutions~\cite{Halder:2025ytq,Urena-Lopez:2005pzi,LinaresCedeno:2021aqk,Roy:2017uvr}. Dynamical systems methods have been further applied to several alternative dark energy scenarios, including Quintom cosmology~\cite{Guo:2004fq,Zhang:2005kj,Lazkoz:2006pa,Leon:2014yua,Roy:2023vxk,Thanankullaphong:2026anl}, k-essence models~\cite{Armendariz-Picon:2000ulo,Piazza:2004df,Chiba:1999ka,Gumjudpai:2005ry,Hussain:2024qrd,Chatterjee:2021ijw}, tachyon and Dirac-Born-Infeld field cosmologies~\cite{Copeland:2010jt,Guo:2008sz,Gumjudpai:2009uy,Gumjudpai:2025tkc,Copeland:2004hq,Quiros:2009mz,Hussain:2022dhp}, as well as bouncing cosmological models~\cite{Maier:2017dtb,Bruni:2021msx,Maier:2013hr,Ye:2026ffe}. These techniques have also played a significant role in the study of modified gravity theories, where they are used to investigate the qualitative evolution of cosmological solutions and the stability of accelerating phases. In particular, phase-space analyzes in $f(R)$ gravity have been extensively explored in~\cite{Starobinsky:1980te,Vinutha:2021ozv,Amendola:2006we,Li:2007xn,Carloni:2015jla,odintsov2017autonomous}. Dynamical systems methods have also been employed in Brans-Dicke theory for both general and specific choices of the Brans--Dicke parameter~\cite{Hrycyna:2013hla,Roy:2017mnz}. Similar approaches have been widely used in scalar-tensor theories of gravity~\cite{faraoni2004scalar,Uzan:1999ch,Gunzig:2000ce,Faraoni:2006sr,Szydlowski:2008in}, where cosmological dynamics often exhibit rich phase-space structures depending on the scalar potential and coupling functions. A broad class of interacting dark energy models has also been investigated using dynamical systems techniques. In particular, interacting quintessence and phantom dark energy scenarios with various choices of scalar-field potentials have been studied in~\cite{Roy:2018eug,Roy:2023uhc,Potting:2021bje,Guo:2004xx,Paliathanasis:2024jxo,Halder:2024aan}. These analyzes provide important insights into the existence and stability of scaling and late-time acceleration solutions. For comprehensive discussions on dynamical systems methods in cosmology and their applications to dark energy and modified gravity theories, we refer the reader to~\cite{coley2003dynamical,Coley:1999uh,Bahamonde:2017ize}.

The field equations of most cosmological models form a set of nonlinear differential equations, for which exact analytical solutions are generally unavailable. The dynamical systems approach provides a powerful qualitative framework for analyzing such nonlinear systems and understanding their global behavior. By introducing suitably normalized, dimensionless variables together with a dimensionless time parameter, the cosmological equations can be recast as an autonomous dynamical system. These variables are typically constructed from physically meaningful quantities and exhibit well-behaved evolution across different cosmological regimes.

Within this framework, fixed points of the dynamical system correspond to asymptotic cosmological solutions, and their stability properties determine the generic evolution of the Universe. In particular, heteroclinic trajectories connecting unstable to stable fixed points naturally describe transitions between different cosmological epochs. Consequently, unstable fixed points may be interpreted as early-time or initial states of the Universe, while stable fixed points represent possible late-time attractors and ultimate cosmological outcomes. 

The purpose of this review is to present a comprehensive and self--contained overview of these alternative dynamical systems methods in cosmology. We begin with a concise introduction to the standard dynamical system formalism widely used in cosmological applications. We then focus on alternative formulations based on polar and hyperbolic transformations, highlighting their conceptual foundations and practical advantages. These methods are subsequently applied to a variety of scalar field dark energy models, with particular emphasis on tracking behavior, initial condition dependence, and global phase space dynamics across a broad range of scalar field potentials. Where appropriate, we refer the reader to detailed technical studies while emphasizing the unifying themes and physical insights that emerge from this line of research.

\section{Basics of Dynamical Systems Analysis}

Here we briefly summarize the essential concepts of dynamical systems analysis relevant for our work; for detailed expositions, see, e.g., Refs.~\cite{hirsch2013differential,perko2013differential,Boehmer:2014vea,Bahamonde:2017ize,Roy:2015jkm}. Dynamical systems theory provides a qualitative framework for analyzing systems of differential equations when exact solutions are unavailable.

\subsubsection{Linear Systems}
An ordinary differential equation (ODE) describes the relation between a function and its derivatives with respect to a single independent variable. We restrict our attention to ODEs, as these naturally arise in cosmological applications. A general system of ODEs can be written as
\begin{equation}
\dot{x} = f(x),
\end{equation}
where $x = (x_1,x_2,\ldots,x_n) \in \mathbb{R}^n$ and $f:\mathbb{R}^n \rightarrow \mathbb{R}^n$. If the system does not explicitly depend on time, it is called an autonomous system. Any non-autonomous system can be converted into an autonomous one by introducing time as an additional variable, thereby increasing the dimensionality of the system by one.

We begin by considering linear systems of the form
\begin{equation}
\dot{x} = Ax,
\end{equation}
where $A$ is a constant matrix $n\times n$. Given an initial condition $x(0)=x_0$, the solution is
\begin{equation}
x(t) = e^{At}x_0,
\end{equation}
where the matrix exponential is defined via its Taylor expansion. The eigenvalues of $A$ determine the qualitative behavior of the solutions. The solution curves of a dynamical system form trajectories in a $n$--dimensional space known as phase space. Of particular importance are fixed (or equilibrium) points, defined by $f(x)=0$, where the system remains stationary. Fixed points are classified according to their stability properties, which are determined by perturbing the system about these points. The existence of a uniqueness theorem for linear systems guaranties the existence of solutions for the linear dynamical system case.

The stability of a fixed point is determined by the eigenvalues of the Jacobian matrix $A$ evaluated at that point. If all eigenvalues have negative real parts, the fixed point is stable and acts as an attractor, meaning that nearby trajectories converge toward it. Conversely, if all eigenvalues have positive real parts, the fixed point is unstable and behaves as a repeller. When the eigenvalues include both positive and negative real parts, the fixed point is classified as a saddle, attracting trajectories along some directions and repelling them along others.

For two-dimensional systems, the qualitative behavior can be visualized directly in phase space, and the classification of fixed points becomes particularly transparent. Let $A$ be a $2 \times 2$ matrix with eigenvalues $\lambda_1$ and $\lambda_2$. The local behavior near the fixed point can then be categorized as follows:
\begin{itemize}
    \item If $\lambda_1 = \lambda_2 < 0$, the fixed point is an attractor focus.
    \item If $\lambda_1 < \lambda_2 < 0$, the fixed point is an attractor node.
    \item If one eigenvalue is negative and the other vanishes, the system exhibits an attracting line.
    \item If $\lambda_1 < 0 < \lambda_2$, the fixed point is a saddle.
    \item If the eigenvalues form a complex conjugate pair $\lambda = a \pm ib$, the nature of the trajectories depends on the sign of $a$: for $a < 0$, trajectories spiral inward (stable focus); for $a > 0$, they spiral outward (unstable focus); and for $a = 0$, the fixed point is a center characterized by closed orbits.
\end{itemize}

More generally, fixed points are classified as hyperbolic or non-hyperbolic. A fixed point is said to be hyperbolic if none of its eigenvalues has zero real part, i.e., $\mathrm{Re}(\lambda_i) \neq 0$ for all $i$. If at least one eigenvalue has vanishing real part, the fixed point is non-hyperbolic, and linear stability analysis alone is insufficient to determine its behavior.

\subsubsection{Non-Linear Systems}
We now turn to the phase space analysis of nonlinear dynamical systems. 
Consider a general autonomous system of differential equations written as
\begin{equation}
\dot{x} = f(x),
\end{equation}
where $f : E \to \mathbb{R}^n$ and $E$ is an open subset of $\mathbb{R}^n$. 
In contrast to linear systems, nonlinear systems cannot, in general, be expressed globally in matrix form. 
However, in the vicinity of a hyperbolic fixed point, the dynamics can be approximated by a linear system.

Let $x^*$ be a fixed point of the system, i.e., $f(x^*) = 0$. 
Introducing a small perturbation $\zeta(t)$ around the fixed point,
\begin{equation}
\zeta(t) = x(t) - x^*,
\end{equation}
we differentiate with respect to time to obtain
\begin{equation}
\dot{\zeta} = \dot{x} = f(x) = f(x^* + \zeta).
\end{equation}

Expanding $f(x^* + \zeta)$ in a Taylor series around $x^*$ gives
\begin{equation}
f(x^* + \zeta) = f(x^*) + Df(x^*)\,\zeta + \mathcal{O}(\zeta^2),
\end{equation}
where $Df(x^*)$ denotes the Jacobian matrix evaluated at the fixed point,
\begin{equation}
Df(x^*) = \left( \frac{\partial f_i}{\partial x_j} \right)_{x=x^*},
\qquad i,j = 1,\dots,n.
\end{equation}

Since $f(x^*)=0$ by definition of the fixed point and $\zeta$ is assumed to be small, higher-order terms can be neglected. 
The perturbation equation therefore reduces to the linear system
\begin{equation}
\dot{\zeta} = Df(x^*)\,\zeta.
\end{equation}

This linear approximation is known as the \emph{linearization} of the nonlinear system near the fixed point. 
The stability properties of the fixed point are determined by the eigenvalues of the Jacobian matrix $A = Df(x^*)$. 
If all eigenvalues have nonzero real parts, the fixed point is hyperbolic and its local behavior is fully characterized by the linearized system.

\paragraph{Existence and Uniqueness Theorem}

A fundamental result in the theory of ordinary differential equations is the existence and uniqueness theorem. 

Let $E$ be an open subset of $\mathbb{R}^n$ containing a point $x_0$, and assume that the vector field $f$ is continuously differentiable on $E$, i.e., $f \in C^1(E)$. Then there exists a positive number $a > 0$ such that the initial value problem
\begin{equation}
\dot{x} = f(x), 
\qquad 
x(0) = x_0,
\end{equation}
admits a unique solution $x(t)$ defined on the interval $[-a, a]$.

This theorem guarantees that, for sufficiently smooth vector fields, the dynamical system generates a well-defined local flow in phase space. In particular, it ensures that trajectories do not intersect and that the time evolution of the system is uniquely determined by its initial conditions.

\paragraph{The Stable Manifold Theorem}

The stable manifold theorem provides a precise description of the structure of trajectories near a hyperbolic fixed point of a nonlinear dynamical system.

Let $E$ be an open subset of $\mathbb{R}^n$ containing the origin, and consider the autonomous system $\dot{x} = f(x),$ with $f \in C^1(E)$. Suppose that $x=0$ is a fixed point, i.e., $f(0)=0$, and let $A = Df(0)$ denote the Jacobian matrix evaluated at the origin. Assume that $A$ has $k$ eigenvalues with negative real parts and $n-k$ eigenvalues with positive real parts.

Then there exists a $k$-dimensional differentiable manifold $W^s(0)$, called the \emph{stable manifold}, which is tangent at the origin to the stable subspace $E^s$ of the linearized system. This manifold is invariant under the flow $\phi_t$ of the nonlinear system. Similarly, there exists an $(n-k)$-dimensional differentiable manifold $W^u(0)$, called the \emph{unstable manifold}, which is tangent at the origin to the unstable subspace $E^u$ of the linearized system. This manifold is also invariant under the flow. The stable and unstable manifolds therefore describe the sets of initial conditions whose trajectories approach the fixed point asymptotically in forward and backward time, respectively.

\paragraph{The Hartman--Grobman Theorem}

The Hartman--Grobman theorem establishes the fundamental connection between a nonlinear system and its linearization in the neighborhood of a hyperbolic fixed point.

Let $E$ be an open subset of $\mathbb{R}^n$ containing the origin, and consider the autonomous system: $\dot{x} = f(x),$ where $f \in C^1(E)$ and $f(0)=0$. Let $\phi_t$ denote the flow generated by this system, and let 
\begin{equation}
A = Df(0)
\end{equation}
be the Jacobian matrix evaluated at the fixed point. Suppose that $A$ has no eigenvalues with a zero real part, so that the fixed point is hyperbolic.

Then there exist open neighborhoods $U$ and $V$ of origin and a homeomorphism $H : U \to V$ such that, for each initial condition $x_0 \in U$, there exists a time interval $I_0 \subset \mathbb{R}$ containing $0$ for which there is a $0$
\begin{equation}
H\big(\phi_t(x_0)\big) = e^{At} H(x_0),
\qquad \text{for all } t \in I_0.
\end{equation}

Near a hyperbolic fixed point, a nonlinear system is topologically equivalent to its linearization $\dot{\zeta} = A\zeta$. The associated homeomorphism preserves the qualitative structure of trajectories, thereby justifying the use of eigenvalue analysis of the Jacobian matrix, similar to that for linear systems, to determine local stability of nonlinear systems. 

However, this approach applies only to hyperbolic equilibria. If a fixed point is non-hyperbolic, i.e., at least one eigenvalue has zero real part, linearization may be inconclusive, and more refined tools such as Lyapunov functions, the Center Manifold Theorem, or numerical methods must be employed.

A commonly encountered situation in cosmological dynamical systems is that of a \emph{normally hyperbolic invariant set}\cite{Coley:1999uh}, where fixed points form a continuous structure such as a curve or surface. Although each point in such a set has at least one zero eigenvalue and is therefore non-hyperbolic, stability can still be determined from the transverse eigenvalues. For instance, if the fixed points form a one-dimensional curve in a $n$-dimensional phase space, stability is governed by the remaining $n-1$ eigenvalues: negative real parts imply attraction, whereas positive real parts indicate repulsion.

\paragraph{The Center Manifold Theorem} The center manifold theorem plays a crucial role in the analysis of non-hyperbolic fixed points, where linear stability analysis alone is insufficient.

Let $E$ be an open subset of $\mathbb{R}^n$ containing the origin, and consider the autonomous system
\begin{equation}
\dot{x} = f(x),
\end{equation}
where $f \in C^r(E)$ with $r \ge 1$. Suppose that $x=0$ is a fixed point, i.e., $f(0)=0$, and let $A = Df(0)$ denote the Jacobian matrix evaluated at the origin. Assume that the spectrum of $A$ consists of: $k$ eigenvalues with negative real parts, $j$ eigenvalues with positive real parts, and $m = n - k - j$ eigenvalues with zero real parts. Then there exist invariant manifolds associated with these spectral subspaces:
\begin{itemize}
    \item a $m$-dimensional \emph{center manifold} $W^c(0)$ of class $C^r$, tangent at the origin to the center subspace $E^c$ of the linearized system;
    \item a $k$-dimensional \emph{stable manifold} $W^s(0)$ of class $C^r$, tangent at the origin to the stable subspace $E^s$;
    \item a $j$-dimensional \emph{unstable manifold} $W^u(0)$ of class $C^r$, tangent at the origin to the unstable subspace $E^u$.
\end{itemize}

Moreover, these manifolds are invariant under the flow $\phi_t$ of the nonlinear system. In particular, the long-term behavior of trajectories near a non-hyperbolic fixed point is determined by the dynamics restricted to the center manifold. Thus, the center manifold theorem reduces the dimensionality of the stability problem by allowing the analysis to focus on the reduced system defined in $W^c(0)$.

\paragraph{Lyapunov Functions}

Determining the stability of non-hyperbolic fixed points is generally more subtle than in the hyperbolic case, since linearization may fail to provide conclusive results. An alternative approach is provided by Lyapunov's direct method, which allows one to infer stability properties without explicitly solving the system.

Let $E$ be an open subset of $\mathbb{R}^n$ containing a fixed point $x_0$, and consider the autonomous system $\dot{x} = f(x),$ where $f \in C^1(E)$ and $f(x_0)=0$. Suppose that there exists a real-valued function $V \in C^1(E)$, known as a \emph{Lyapunov function}, satisfying $V(x_0)=0$ and $V(x)>0 \quad \text{for all } x \neq x_0$.

The stability of $x_0$ can then be determined from the sign of the derivative of $V$ along trajectories,
\begin{equation}
\dot{V}(x) = \nabla V \cdot f(x).
\end{equation}

\begin{itemize}
    \item If $\dot{V}(x) \le 0$ for all $x \in E$, the fixed point $x_0$ is stable.
    \item If $\dot{V}(x) < 0$ for all $x \in E \setminus \{x_0\}$, the fixed point is asymptotically stable.
    \item If $\dot{V}(x) > 0$ in a neighborhood of $x_0$, the fixed point is unstable.
\end{itemize}

Lyapunov functions therefore provide a powerful tool for establishing stability properties in cases where eigenvalue analysis is inconclusive, particularly for non-hyperbolic fixed points encountered in cosmological dynamical systems.

\section{ Dynamical Systems Formalism in Cosmology}
In this section, we briefly review the  dynamical systems approach and its application to cosmological studies.
The equations governing most cosmological models form a system of nonlinear differential equations, for which exact analytical solutions are often unavailable. The dynamical systems approach provides a powerful framework for extracting the qualitative behavior of such models by introducing suitably normalized dimensionless variables and a dimensionless time parameter, thereby recasting the equations into an autonomous system. These variables are typically constructed from physically meaningful quantities and remain well behaved throughout the cosmic evolution. By identifying the fixed points of the system and analyzing their stability properties, one can characterize different cosmological regimes and infer the generic past and future behavior of the Universe. In particular, heteroclinic trajectories connecting unstable to stable fixed points naturally describe transitions between cosmological epochs, suggesting that unstable equilibria may represent early-time states, while stable attractors determine the asymptotic fate of the Universe. This methodology has been widely applied in general relativity and cosmology, including studies of modified gravity theories, scalar–tensor models, Bianchi cosmologies, and non-minimally coupled scalar field scenarios\cite{Boehmer:2014vea,Bahamonde:2017ize,Roy:2015jkm, Coley:1999uh}.

\subsection{Scalar Field Cosmology} 
For example, to illustrate how dynamical-system methods can be applied in scalar field cosmology, we begin with a scalar field as a candidate for dark energy. We work in a spatially flat, homogeneous, and isotropic Friedmann--Lemaître--Robertson--Walker (FLRW) Universe, described by the metric
\begin{equation}
 ds^{2} = -dt^{2} + a^{2}(t)\left(dx^{2} + dy^{2} + dz^{2}\right),
\end{equation}
with $a(t)$ being the scale factor. 

The matter content consists of a minimally coupled scalar field $\phi$ and a perfect fluid that accounts for ordinary matter or radiation. The evolution of the system is then determined by the Friedmann equations together with the scalar-field equation of motion,
\begin{align}\label{eq:quint}
H^{2} &= \frac{\kappa^{2}}{3}
\left( \rho_\mathrm{b} + \rho_{\phi} \right),  \\
\dot{H} &= -\frac{\kappa^{2}}{2}
\left[ \left( \rho_\mathrm{b} + p_\mathrm{b} \right)
+ \left( \rho_{\phi} + p_{\phi} \right) \right],  \\
\dot{\rho}_{j} &= -3H \left( \rho_\mathrm{b} + p_\mathrm{b} \right),  \\
\ddot{\phi} &= -3H \dot{\phi} - \epsilon \frac{dV(\phi)}{d\phi}, \label{eq:quint}
\end{align}
where $\kappa^{2} = 8\pi G$, $H \equiv \dot{a}/a$ is the Hubble parameter and $\rho_\mathrm{b}$ and $p_\mathrm{b}$ denote the density and pressure of the perfect fluid, respectively. The function $V(\phi)$ represents the scalar field potential, and an overdot denotes differentiation with respect to cosmic time. Here we introduce the switch parameter $\epsilon$ such that $\epsilon = +1$ corresponds to quintessence field and $\epsilon = -1$ corresponds to phantom field. The equations of motion can be generalized to include many perfect fluids, but for simplicity we will restrict ourselves to one background fluid apart from the scalar field.

The energy density $\rho_{\phi}$ and pressure $p_{\phi}$ associated with the scalar field are given by
\begin{equation}
\rho_{\phi} = \epsilon \frac{1}{2}\dot{\phi}^{2} + V(\phi),
\qquad
p_{\phi} = \epsilon \frac{1}{2}\dot{\phi}^{2} - V(\phi).
\label{eq:KG_quint}
\end{equation}

To study the dynamics of scalar field models, one may solve the Klein–Gordon equation~(\ref{eq:KG_quint}) together with the Einstein field equations~(\ref{eq:quint}), either analytically or numerically. However, these equations form a system of coupled second-order differential equations, whose direct solution is often technically demanding. A more efficient approach is to introduce suitable dimensionless variables that recast the system particularly the Klein–Gordon equation into a set of first-order autonomous differential equations. This transformation significantly simplifies the analysis and enables the application of dynamical system techniques to investigate the qualitative behavior of cosmological evolution.

\subsection{Autonomous System: the standard transformation}

To recast the cosmological equations into an autonomous dynamical system, it is customary to use dimensionless variables normalized by the Hubble scale, first introduced in the seminal paper~\cite{Copeland:1997et},
\begin{equation} \label{eq:d_sys_original}
x \equiv \frac{\dot{\phi}}{\sqrt{6}H}, \qquad
y \equiv \frac{\sqrt{V(\phi)}}{\sqrt{3}H}.
\end{equation}
In terms of these variables, and using Eqs.~\eqref{eq:quint}, the scalar field density parameter $\Omega_{\phi}$ becomes
\begin{equation}
\Omega_{\phi} \equiv \frac{\kappa^2 \rho_\phi}{3H^2} = \epsilon x^{2} + y^{2} \, ,
\end{equation}
while the scalar field equation of state parameter $w_\phi$ is
\begin{equation}
w_{\phi} = \frac{p_\phi}{\rho_\phi} = \frac{\epsilon x^{2} - y^{2}}{\epsilon x^{2} + y^{2}} \, .
\end{equation}
Likewise, from the Friedmann constraint we obtain the following,
\begin{equation}
    1 = \frac{\kappa^2 \rho_\mathrm{b}}{3H^2} + \frac{\kappa^2 \rho_\phi}{3H^2} = \Omega_\mathrm{b} + \epsilon x^2 +y^2 \, ,
\end{equation}
then the density parameter of the background fluid $\Omega_\mathrm{b}$ can also be written as $\Omega_\mathrm{b} = 1 - \epsilon x^2 -y^2$.

Using the logarithmic variable $N \equiv \ln (a/a_i)$, where $a_i$ is a chosen initial value for the scale factor, the evolution equations can be written as an autonomous system,
\begin{align}
x' &= -3x + \epsilon \lambda \sqrt{\frac{3}{2}}\, y^{2}
+ \frac{3}{2}x\left[2 \epsilon x^{2} + \gamma_\mathrm{b} \left(1 - \epsilon x^{2} - y^{2}\right)\right] \, , \\
y' &= -\lambda \sqrt{\frac{3}{2}}\, x y
+ \frac{3}{2}y\left[2 \epsilon x^{2} + \gamma_\mathrm{b} \left(1 - \epsilon x^{2} - y^{2}\right)\right] \, , \\
\lambda' &= -\sqrt{6}\,\lambda^{2}\,(\Gamma - 1)\,x \, ,  
\end{align}
where a prime denotes the derivative with respect to $N$ and $\gamma_\mathrm{b} = p_\mathrm{b}/\rho_\mathrm{b}$ is the equation of state parameter of the background fluid. The new dimensionless parameters $\lambda$ and $\Gamma$ are given by
\begin{equation}
\lambda \equiv -\frac{1}{\kappa V}\frac{dV}{d\phi} \, , \quad \Gamma = \frac{V\,\dfrac{d^{2}V}{d\phi^{2}}}{\left(\dfrac{dV}{d\phi}\right)^{2}} \, , \label{eq:lambda}
\end{equation}
which characterize the properties of the chosen scalar field potential $V(\phi)$. 

The simplest possibility is to consider $\lambda = \mathrm{const.}$, which leads to the exponential potential $V(\phi) = V_0 e^{-\lambda \kappa \phi}$ studied in~\cite{Copeland:1997et} and also to $\Gamma = 1$, and in consequence the system nicely closes as a two--dimensional autonomous dynamical system for the variables $x$ and $y$ only. Another possibility of closing the system of equations is to express $\Gamma$ as a function of $\lambda$, see, for instance~\cite{Roy:2017uvr}, where a comprehensive list of forms of $\Gamma (\lambda)$ for different choices of field parameters in $V(\phi)$. For more general potentials, both parameters become field-dependent, $\lambda = \lambda(\phi)$ and $\Gamma = \Gamma (\phi)$, and the system may have to be extended.


\subsection{Autonomous System: the polar transformation}
Although the standard formulation $(x,y)$ has proven extremely successful, the geometric interpretation of the phase space is not always transparent. In particular, the scalar field energy density and its equation of state are encoded in nonlinear combinations of the dynamical variables, which may obscure the physical meaning of trajectories. Moreover, cosmological solutions of interest do not always appear in a geometrically simple form in the $(x,y)$ plane. This can make the analysis of initial conditions and asymptotic regimes less intuitive.

To have an alternative form of the autonomous system using Eqs.~(\ref{eq:d_sys_original}) together with Eqs.~(\ref{eq:KG_quint}), we introduce a new set of dimensionless variables.
\begin{align} \label{eq:DS_xy}
x &= \frac{\kappa \dot{\phi}}{\sqrt{6}H} = \Omega_\phi ^{1/2} \sin (\theta /2) \, , 
\qquad
y =\frac{\kappa \sqrt{V}}{\sqrt{3}H} = \Omega_\phi ^{1/2} \cos (\theta /2) \, ,  \\
y_{1} &= -\frac{2\sqrt{2}\,\partial_{\phi}\sqrt{V}}{H},
\qquad
y_{2} = -\frac{4\sqrt{3}\,\partial_{\phi}^{2}\sqrt{V}}{\kappa H}. 
\end{align}
where $\Omega_\phi$ is again the scalar field energy density, bounded in the range $[0,1]$, whereas the new angular variable $\theta \in [0,2\pi)$. The equation of state (EoS) of the scalar field in terms of the polar variable is given by
\begin{equation}
    w_{\phi} = \frac{p_{\phi}}{\rho_{\phi}} = \frac{x^{2} - y^{2}}{x^{2} + y^{2}} = -\cos\theta \label{eq:polar_EoS}
\end{equation} 
which shows the direct relation between the two variables. 

Using the polar transformation given above in Eq.(\ref{eq:DS_xy}), the autonomous system can be transformed into an autonomous system, and the resulting equations take the general form
\begin{subequations}\label{eq:polar}
\begin{align}
\theta' &= -3\sin\theta + y_{1}, \label{eq:polar_th}\\
y_{1}' &= \frac{3}{2} \gamma_\mathrm{tot} \, y_{1} + \Omega_{\phi}^{1/2}\sin(\theta/2)\,y_{2}, \label{eq:polar_y1} \\
\Omega_{\phi}' &= 3 (\gamma_\mathrm{tot} - \gamma_\phi ) \Omega_{\phi} \, , \label{eq:polar_om}
\end{align}
\end{subequations}
where for convenience we are using $\gamma_\phi = 1 - \cos \theta$, to denote the equations of state of the scalar field, and also that
\begin{equation}
    \gamma_\mathrm{tot} = 1 + \frac{p_{\mathrm{tot}}}{\rho_{\mathrm{tot}}} = \gamma_\mathrm{b} - ( \gamma_\mathrm{b} - \gamma_\phi ) \, \Omega_\phi \, ,
\end{equation}

The difference in this formalism is the introduction of the new dimensionless variables $y_1$ and $y_2$ compared to the $\lambda$ and $\Gamma$ variables in the standard approach presented above. This polar transformation was first used in \cite{Urena-Lopez:2015odd} and later extensively used to study both the scalar field dark matter \cite{Urena-Lopez:2015gur, Cedeno:2017sou,Urena-Lopez:2023ngt} and dark energy scenarios \cite{Roy:2018nce,Roy:2018eug,Urena-Lopez:2020npg,LinaresCedeno:2021aqk,Roy:2023vxk} both at the background and perturbation level.

The equations of motion under the polar transformation are the same for the scalar potential $V(\phi)$, and only the equation for $y_1$ changes for different cases due to the presence of the new variable $y_{2}$. In analogy to the standard dynamical systems approach, the potential-dependent variables $y_1$ and $y_2$ must, in general, be computed explicitly for each specific choice of quintessence potential according to their definitions. This model-by-model treatment can limit the generality of the formalism. 

However, in \cite{Roy:2018nce} it was shown that a unified parameterization of the variable $y_2$ can be introduced in the following polynomial form
\begin{equation}
y_2 = y \sum_{i=0}^{n} \alpha_i \left( \frac{y_1}{y} \right)^i  \, , \label{eq:y2-polynomial}
\end{equation}
where constant coefficients $\alpha_i$ characterize the underlying scalar field potential. Note that $y_1/y = \sqrt{6} \lambda$, with $\lambda$ as defined in Eq.~\eqref{eq:lambda}, and then Eq.~\eqref{eq:y2-polynomial} can also be rewritten in polynomial form in terms of $\lambda$.

Remarkably, already at $n=2$, this parametrization ($y_2 = \alpha_0 y + \alpha_1 y_1+ \alpha_2 {y_1}^2 /y$) is sufficiently general to reproduce a large class of commonly studied potentials within a single framework.  The proposed form was obtained by explicitly evaluating $y_2$ for a broad range of potentials considered in the literature and identifying a common functional structure. Furthermore, the approach is constructive: given a particular choice of the coefficients $\alpha_i$, one can recover the corresponding scalar field potential by integrating the defining relations in reverse. In this way, the parametrization not only unifies many known models but also generates new classes of potentials in a systematic manner. The general forms of these potentials are summarized in Table \ref{tab:potentials}. This framework encompasses a broad range of widely used potentials found in the literature (see Table II of \cite{Roy:2018nce}).

\begin{table}[h!]
\small
\centering
\caption{Catalogue of generic quintessence potentials derived by reverse-integrating the definition of the second potential variable $y_2$. The expansion is truncated at second order. \label{tab:potentials}}
\resizebox{\textwidth}{!}{%
\begin{tabular}{|c|c|c|}
\hline
No & Structure of $y_2/y$ & General form of the potentials $V(\phi)$  \\
\hline
Ia & $\alpha_0=0,\;\alpha_1=0,\;\alpha_2=-\tfrac{1}{2}$  & 
$(A + B\phi)^{\frac{2}{2\alpha_2 + 1}}$ \\
\hline
Ib & $\alpha_0=0,\;\alpha_1=0,\;\alpha_2=-\tfrac{1}{2}$ &
$A^2 e^{2B\phi}$ \\
\hline
IIa & $\alpha_0=0,\;\alpha_1=0,\;\alpha_2=-\tfrac{1}{2}$ &
$ A^2 \cos\!\left( \frac{\sqrt{\alpha_0 \kappa^2 (1 + 2\alpha_2)} \, (\phi - B)}{2\sqrt{3}} \right)^{\tfrac{2}{1 + 2\alpha_2}}$ \\
\hline
IIb & $\alpha_0=0,\;\alpha_1=0,\;\alpha_2=-\tfrac{1}{2}$ &
$A^2 \exp\!\left(-\tfrac{\kappa^2\alpha_0}{12}\phi^2\right)\exp(2B\phi)$  \\
\hline
IIIa & $\alpha_0=0,\;\alpha_1=0,\;\alpha_2=-\tfrac{1}{2}$ &
$A \exp\!\left(\tfrac{\alpha_1\kappa\phi}{\sqrt{6}}+B\right)^{\tfrac{2}{1+2\alpha_2}}$  \\
\hline
IIIb & $\alpha_0=0,\;\alpha_1=0,\;\alpha_2=-\tfrac{1}{2}$ &
$A^2 \exp\!\left(2B \exp\!\left(\tfrac{\kappa\alpha_1\phi}{\sqrt{6}}\right)\right)$  \\
\hline
IVa & $\alpha_0=0,\;\alpha_1=0,\;\alpha_2=-\tfrac{1}{2}$ &
$A^2 \exp\!\left(\tfrac{\kappa\alpha_1\phi}{\sqrt{6}(1+2\alpha_2)}\right)
\cos\!\left(-\tfrac{\kappa^2\alpha_2}{12}+\tfrac{\kappa^2\alpha_0}{12(1+2\alpha_2)}\right)^{\tfrac{1}{2}}(\phi-B)^{\tfrac{2}{1+2\alpha_2}}$  \\
\hline
IVb & $\alpha_0=0,\;\alpha_1=0,\;\alpha_2=-\tfrac{1}{2}$ &
$A^2 \exp\!\left(\tfrac{\kappa\alpha_0\phi}{\sqrt{6}\alpha_1}\right)
+2B \exp\!\left(\tfrac{\kappa\alpha_1\phi}{\sqrt{6}}\right)$ \\
\hline
\end{tabular}}
\end{table}

\subsection{Critical Points and Physical Interpretation}
Qualitative cosmological behavior is determined by the critical points of the autonomous system. Each critical point corresponds to a distinct cosmological solution, and to determine whether a given critical point can describe the early- or late-time Universe, one performs a linear stability analysis. 


The critical points of the polar dynamical system satisfy $\Omega'_\phi = \theta' = y'_1 = 0$, under which we shall obtain the values $(\Omega_{\phi c}, \theta_c, y_{1c})$. Imposing the condition $\theta' = 0$ (see Eq.~(\ref{eq:polar_th})) immediately yields
\begin{equation}
y_{1c} = 3 \sin\theta_c,
\end{equation}
where the subscript $c$ denotes the evaluation at the critical point. Substituting this result into Eq.~(\ref{eq:polar_y1}) and Eq.~(\ref{eq:polar_om}), together with the general expression for $y_2$ given in Eq.~(\ref{eq:y2-polynomial}), we obtain the remaining conditions for the existence of critical points:
\begin{subequations}
    \begin{align}
\left(
\gamma_{\text{tot}}
+ \frac{\alpha_0}{9}\Omega_{\phi c}
+ \frac{\sqrt{2}}{3}\alpha_1 \Omega_{\phi c}^{1/2}\gamma_{\phi c}^{1/2}
+ 2\alpha_2 \gamma_{\phi c}
\right)\sin\theta_c &= 0, \label{eq:9a} \\
(\gamma_\mathrm{tot} - \gamma_{\phi c} ) \Omega_{\phi c} &= 0 \, . \label{eq:9b}
\end{align}
\end{subequations}
In deriving Eq.~(\ref{eq:9b}), we have used the relation $\gamma_{\phi c} = 2 \sin^2\left(\theta_c/2\right),$ and restricted the angular variable to the interval $0 \leq \theta_c \leq \pi$. 

We now summarize the critical points obtained from Eqs.~(\ref{eq:9b}) for scalar field potentials that admit the parameterization given in Eq.~(\ref{eq:y2-polynomial}). In doing so, we adopt the standard classification commonly used in the literature.

\paragraph{(a) Fluid-dominated solution.}
This case corresponds to $\Omega_{\phi c}=0$ in Eq.~(\ref{eq:9b}), implying that the contribution of the scalar field to the total energy density is generally negligible. Note that the value $\Omega_{\phi}=0$ is an exact solution of Eq.~\eqref{eq:polar_om}, and the scalar field would never contribute to the cosmic density under such an initial condition. In practice, one usually considers $\Omega_\phi \ll 1$ such that the scalar field starts its evolution very closely to the critical point of fluid domination $\Omega_{\phi c}=0$ at early times.

From Eq.~(\ref{eq:9a}), consistency then requires $\theta_c = 0$ or $\pi$, which leads to $\gamma_{\phi c}=0$ or $2$, respectively. For $\gamma_{\phi c}=0$ the evolution of the scalar field is indistinguishable from the cosmological constant, whereas for $\gamma_{\phi c}=2$ the scalar field behaves as a stiff fluid.

\paragraph{(b) Scaling solution.}
An alternative possibility arising from Eq.~(\ref{eq:9b}) with $\Omega_{\phi c} \neq 0$ is $\gamma_{\text{tot}}=\gamma_{\phi c}$, which means that the scalar field equation of state matches that of the dominant background component. The scaling solution is then characterized by a constant ratio between the scalar field energy density and that of the background fluid, that is, $\rho_\phi / \rho_\mathrm{b} = \text{constant}$. The existence of scaling solutions depends on the structure of the potential. 

In general, Eq.~(\ref{eq:9a}) reduces to

\begin{equation} \label{eq:quint_sclaing}
\alpha_0 \Omega_{\phi c}
+ 3\sqrt{2}\,\alpha_1 \Omega_{\phi c}^{1/2}\gamma_{\text{tot}}^{1/2} + 9(1+2\alpha_2)\gamma_{\text{tot}} = 0 \, .
\end{equation}
Equation~(\ref{eq:quint_sclaing}) allows for extended scaling solutions when the so-called active parameters $\alpha_i$ are nonzero. However, a physically meaningful scaling solution for $\Omega_{\phi c}$ exists only for specific combinations of the parameters $\alpha_i$, such that the expression above admits real solutions with $0<\Omega_{\phi c}<1$. In the particular case $\alpha_1=0=\alpha_2$, one recovers the standard scaling result
\begin{equation}
\Omega_{\phi c} = -\frac{9\gamma_{\text{tot}}}{\alpha_0},
\end{equation}
which requires $\alpha_0<0$, which means that this scaling solution corresponds to the type IIa potential in Table~\ref{tab:potentials}.

Exponential potentials of the form $V(\phi) = V_0 e^{-\lambda \kappa \phi}$ admit exact scaling fixed points for a constant $\lambda$, as explained in~\cite{Copeland:1997et}. However, note that according again to Table~\ref{tab:potentials}, an exponential potential is a type Ib potential in Table~\ref{tab:potentials}, for which the active parameters are $\alpha_0 =0$, $\alpha_1=0$ and $\alpha_2 = -1/2$, and then the scaling condition~\eqref{eq:quint_sclaing} is trivially satisfied.

Although scaling solutions prevent early domination of the scalar field, they do not lead to cosmic acceleration by themselves. Therefore, in the case of dark energy, additional mechanisms are required for late-time acceleration of the expansion of the universe.

\paragraph{(c) Scalar field-dominated solution.}
This regime is characterized by $\Omega_{\phi c}=1$ and $\gamma_{\text{tot}}=\gamma_{\phi c}$. Under these conditions, Eq.~(\ref{eq:9a}) becomes
\begin{equation} \label{eq:quint_domination}
\left[
\alpha_0
+ 3\sqrt{2}\,\alpha_1 \gamma_{\phi c}^{1/2}
+ 9(1+2\alpha_2)\gamma_{\phi c}
\right]\sin\theta_c = 0.
\end{equation}

The simplest solutions correspond to $\theta_c=0$ or $\pi$, giving $\gamma_{\phi c}=0$ or $2$, respectively. These cases describe scalar field domination by either the potential energy $V(\phi)$ or the kinetic term $\dot{\phi}^2/2$. Additional solutions may arise from the vanishing of the expression inside the brackets in Eq.~(\ref{eq:quint_domination}). Such solutions depend on the specific values of the active parameters $\alpha_i$ and, for $0<\gamma_{\phi c}<2$, correspond to scalar field-dominated epochs with a well-defined equation of state determined by these parameters.

\paragraph{(d)Tracker solution.}
One particularly interesting benefit of this alternative approach is that one can easily identify a general class of tracker solutions for the quintessence model. A tracker is an attractor-like solution toward which a wide range of initial conditions converge during cosmic evolution. Unlike pure scaling solutions, tracker solutions do not necessarily satisfy $w_\phi = w_\mathrm{b}$ at all times; instead, the scalar field dynamically adjusts its equation of state and eventually evolves toward domination.

For example, in the class type I of potentials for which $\alpha_0 = 0 = \alpha_1$, see Table~\ref{tab:potentials}, Eq.~(\ref{eq:9a}) yields the critical condition
\begin{equation} \label{eq:tracker_quint}
\gamma_{\phi c} = -\frac{\gamma_{\text{tot}}}{2\alpha_2},
\end{equation}
which corresponds to the tracker condition. Since a physically meaningful equation of state requires $0 < \gamma_{\phi c} < 2$, the above relation implies $0 < -\frac{\gamma_{\text{tot}}}{\alpha_2} < 4,$, which leads to the constraint $\alpha_2 < -\gamma_{\text{tot}}/4$. If the tracker regime is to be applied during radiation domination, where $\gamma_{\text{tot}} = 4/3$, this condition reduces to $\alpha_2 < -\frac{1}{3}$.

In addition, tracker solutions with $\gamma_{\phi c} < \gamma_{\text{tot}}$ are phenomenologically favored, as they allow the scalar field equation of state to approach $w_\phi \rightarrow -1$ at late times. As an illustrative example, consider power-law quintessence models of the form $V(\phi) = M^{4-p}\phi^{p},$ for which $p = 2/(1+2\alpha_2)$ when $\alpha_0=\alpha_1=0$. In this case, the tracker condition~(\ref{eq:tracker_quint}) becomes $\gamma_{\phi c} = \frac{p}{p-2} \,\gamma_{\text{tot}}$. The existence of a tracker solution requires $p>6$ (corresponding to $-1/2<\alpha_2<-1/3$) or $p<0$ (for $\alpha_2<-1/2$). The special case $\alpha_2=-1/2$ corresponds to an exponential potential.



\subsection{Determination of Initial Conditions}
Since we are interested in implementing this model in the Boltzmann code Cosmic Linear Anisotropy Solving System (\texttt{CLASS}) \footnote{A recent Python implementation of dynamical-systems-based methods for cosmological data analysis at the background level is described in \cite{Roy:2026icy}. This code offers a framework for constraining cosmological models within a dynamical system formulation and is integrated with the Cobaya interface, enabling users to exploit the advanced statistical and inference capabilities already available in Cobaya for cosmological parameter estimation and model analysis. The code is publicly accessible at \href{https://github.com/Nandancosmos/CosmoDS}{https://github.com/Nandancosmos/CosmoDS}.}~\cite{Lesgourgues:2011rh,Blas:2011rf} in this section, we outline a practical procedure for estimating the initial conditions of thawing and tracker quintessence models within the polar dynamical systems framework. Our goal is to obtain approximate analytical expressions that are sufficiently accurate to initialize numerical integrations. The reader should keep in mind that these are not exact solutions, but rather an approximation scheme for the better performance of the numerical integration.

\paragraph{(a) Thawing quintessence.}
One of the core approximations of this initial condition estimation is that the dark energy is thawing in nature. The thawing nature of the dark energy is characterized by an equation of state which at an early time is close to that of a cosmological constant $w_\phi \simeq -1,$ and differs from it at a late time which in polar variables corresponds to $\theta \ll 1$. In addition, during radiation- and matter-dominated epochs, the scalar field energy density remains subdominant $\Omega_\phi \ll 1$. Under these conditions, the dynamical equations simplify considerably, allowing analytical solutions in both eras.

During radiation domination, the total equation of state is $w_{\text{tot}} = 1/3$. Using $\sin\theta \simeq \theta$ and $\cos\theta \simeq 1$, the evolution equations reduce to $\theta' = -3\theta + y_1, \quad$ $y_1' = 2 y_1, \quad$ and $\Omega_\phi' = 4 \Omega_\phi.$ One can solve these equations and considering the growth mode solutions, one obtains: 
\begin{equation}
 \theta_r = \theta_i \left(\frac{a}{a_i}\right)^2,  \quad y_{1r} = y_{1i} \left(\frac{a}{a_i}\right)^2,  \quad \Omega_{\phi r} = \Omega_{\phi i} \left(\frac{a}{a_i}\right)^4,   
\end{equation}
where the subscript $i$ denotes initial values deep in the radiation era. For example, in the case of the scale factor, one typically has $a_i \sim 10^{-14}$. The attractor relationship in this regime gives us $y_1 = 5\theta$.

Adopting a similar approach to the radiation domination era in the matter-dominated epoch, $w_{\text{tot}} = 0$, and under the same small-angle and subdominant-field approximations, the equations become $\theta' = -3\theta + y_1, \quad y_1' = \frac{3}{2} y_1, \quad \Omega_\phi' = 3 \Omega_\phi.$ Solving these equations yields
\begin{align*}
\theta_m = 
\left(\theta_{eq} - \frac{2}{9}y_{1eq}\right)
\left(\frac{a}{a_{eq}}\right)^{-3}
+ \frac{2}{9}y_{1eq}
\left(\frac{a}{a_{eq}}\right)^{3/2}, \quad
y_{1m} = y_{1eq}\left(\frac{a}{a_{eq}}\right)^{3/2}, \quad
\Omega_{\phi m} &= \Omega_{\phi eq}
\left(\frac{a}{a_{eq}}\right)^3,
\end{align*}
where the subscript $eq$ denotes the quantities evaluated with respect to the equality of radiation and matter. To obtain a continuous evolution from the radiation to the matter evolution eras, one can match the radiation- and matter-dominated solutions at $a=a_{eq}$. 




Assuming that the matter-dominated solution remains a good approximation up to the present epoch ($a=1$), one can invert the analytical solutions to estimate the primordial quantities from present-day observables. Keeping only the dominant growing mode, one finds approximately
\begin{equation}
\theta_i \simeq 
\frac{9}{10} a_i^2 
\left(\frac{\Omega_{m0}}{\Omega_{r0}}\right)^{1/2}
\theta_0, \quad \Omega_{\phi i} \simeq A \, a_i^4 \frac{\Omega_{m0}}{\Omega_{r0}} \Omega_{\phi 0},
\end{equation}
where $\Omega_{r0}$, $\Omega_{m0}$, and $\Omega_{\phi 0}$ denote the current density parameters of radiation, matter, and quintessence. The initial value of the variable $y_1$ can be obtained from the condition $y_{1i} = 5\theta_i$ mentioned before, and the constant $A$ represents a free normalization parameter that will be adjusted numerically.






\paragraph{(b) Tracker quintessence.}
Finding the tracker solutions can be very useful in reducing the initial condition sensitivity of the scalar field models. Here we describe how one can find the initial conditions of the dynamical system variables using the tracker condition obtained in Eq~(\ref{eq:tracker_quint}). Using the tracker condition, the initial values of the dynamical variables can be written as
\begin{align}\label{eq:track_quint}
\cos\theta_i &= 1 + \frac{2}{3\alpha_2}, 
\qquad 
y_{1i} = 3\sin\theta_i, 
\\
\Omega_{\phi i} &= A\, a_i^{4\left(1+\frac{1}{2\alpha_2}\right)}
\left(\frac{\Omega_{m0}}{\Omega_{r0}}\right)^{1+\frac{1}{2\alpha_2}}
\Omega_{\phi 0},
\end{align}
where again $\Omega_{r0}$, $\Omega_{m0}$, and $\Omega_{\phi 0}$ denote the current density parameters of radiation, matter, and quintessence, respectively. The expressions for $\theta_i$ and $y_{1i}$ follow directly from the tracker relation, whereas the expression for $\Omega_{\phi i}$ is obtained by integrating Eq.~(\ref{eq:polar_om}) up to the present epoch, assuming tracker evolution during both radiation- and matter-dominated eras. As before, the constant $A$ represents a free normalization parameter that should be adjusted numerically.

\begin{figure}[h!]
    \begin{minipage}[b]{0.55\textwidth}
        \centering \includegraphics[width=\linewidth]{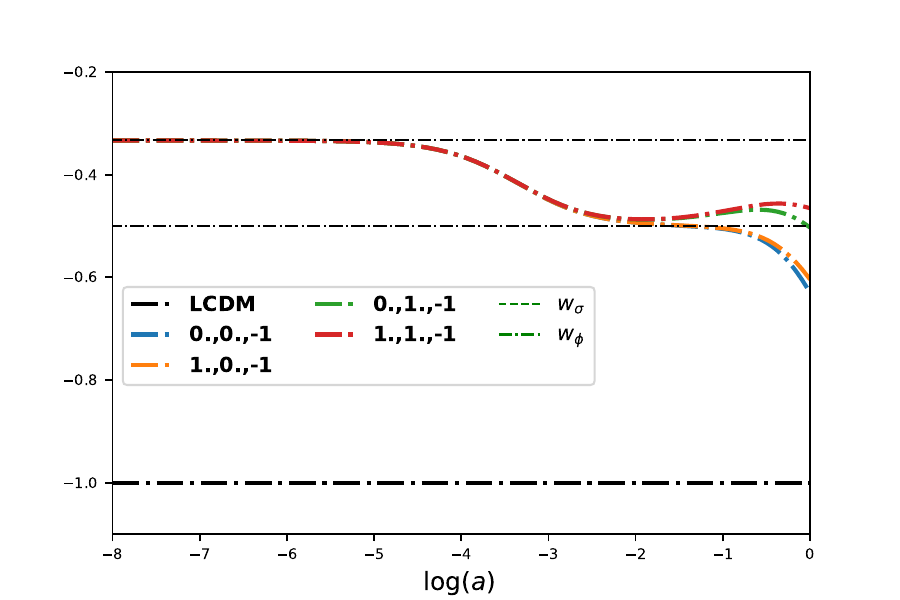}
    \end{minipage}
    \hfill
    \begin{minipage}[b]{0.55\textwidth}
        \centering \includegraphics[width=\linewidth]{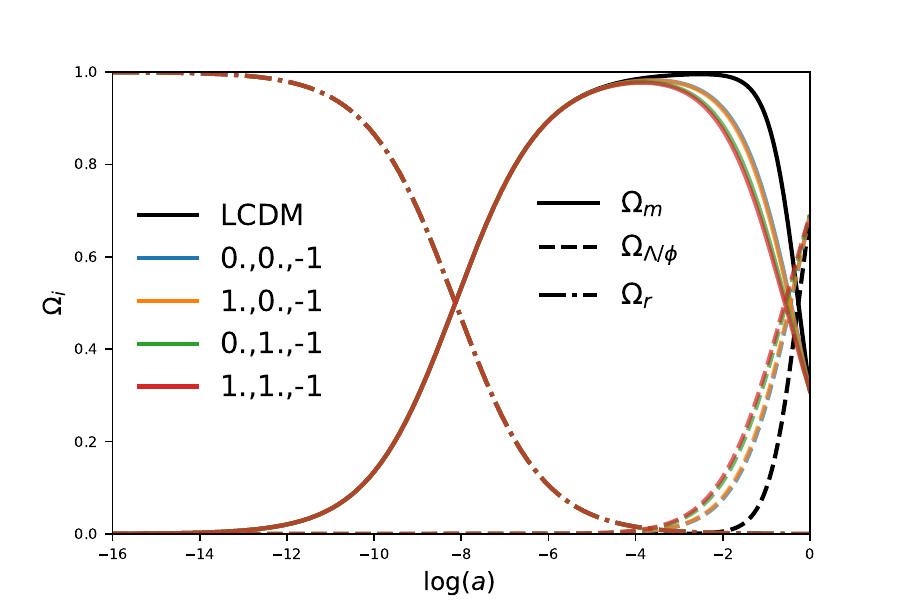}
    \end{minipage}
    \caption{(Left) The evolution of the EoS of the quintessence field with the choice of the tracker initial condition for the corresponding $\alpha_i$ parameter triplets shown in the labels of the plot. The lines in blue are shown for the tracker value of the EoS during matter and radiation domination. (Right) Evolution of the density parameters of the universe for the tracker quintessence for same choice of the $\alpha_i$ parameters.  }\label{fig:quint1}
\end{figure}

These approximated initial conditions can greatly enhance the performance and stability of the Boltzmann codes, for example \textit{CLASS}. In Fig.\ref{fig:quint1} (right) we have shown the evolution of the EoS of the scalar field while using the tracker condition given in Eq.(\ref{eq:track_quint}). The black dotted dashed lines show the tracker value of the EOS during matter and radiation domination. On the right, we have shown the evolution of the density parameters for the tracker solutions.

\subsection{Linear Density Perturbations}

This polar formulation can also be employed to investigate the evolution of linear density perturbations of the scalar field. Let us examine linear perturbations of the scalar field in relation to the homogeneous background solution. We decompose the field as
\begin{equation}
\phi(x,t) = \phi(t) + \varphi(x,t),
\end{equation}
where $\varphi(x,t)$ represents a small perturbation. For the metric, we adopt the synchronous gauge, in which the line element is written as
\begin{equation}
ds^2 = -dt^2 + a^2(t)\left(\delta_{ij} + h_{ij}\right)dx^i dx^j,
\end{equation}
with $h_{ij}$ denoting the metric perturbations. In Fourier space, the perturbation $\varphi(k,t)$ satisfies the linearized Klein--Gordon equation,
\begin{equation}
\ddot{\varphi}
= -3H\dot{\varphi}
-\left(\frac{k^2}{a^2} + V_{,\phi\phi}\right)\varphi
-\frac{1}{2}\dot{h}\dot{\phi},
\end{equation}
where $k$ is the comoving wave number and $h = h^i_{\ i}$.

To connect the perturbation dynamics with the background evolution, it is convenient to introduce suitable dimensionless variables. In analogy with the background polar parametrization, the perturbation variables can be recast into a dynamical system form by using the following transformations:
\begin{equation}
\sqrt{\frac{2}{3}}\,\frac{\kappa \dot{\varphi}}{H}
= -\Omega_{\phi}^{1/2} e^{\beta} \cos\left(\frac{\vartheta}{2}\right),
\qquad
\frac{\kappa y_1 \varphi}{\sqrt{6}}
= -\Omega_{\phi}^{1/2} e^{\beta} \sin\left(\frac{\vartheta}{2}\right).
\end{equation}
Defining the fractional density contrast variables $\delta_0 = -e^{\beta}\sin\!\left(\frac{\theta}{2} - \frac{\vartheta}{2}\right) $ and $\delta_1 = -e^{\beta}\cos\!\left(\frac{\theta}{2} - \frac{\vartheta}{2}\right).$
 associated with the scalar field perturbations, one obtains the coupled system
\begin{align}
\delta_0' &= 
\left[3\sin\theta + \frac{k^2}{k_J^2}(1-\cos\theta)\right]\delta_1
+ \frac{k^2}{k_J^2}\sin\theta\,\delta_0
-\frac{h'}{2}(1-\cos\theta), \\
\delta_1' &= 
-\left(3\cos\theta + \frac{k_{\rm eff}^2}{k_J^2}\sin\theta\right)\delta_1
+ \frac{k_{\rm eff}^2}{k_J^2}(1+\cos\theta)\delta_0
-\frac{h'}{2}\sin\theta,
\end{align}
where a prime denotes differentiation with respect to the number of $e$-folds $N=\ln (a/a_i)$. It can be easily shown that the standard density contrast of density perturbations of the scalar field is given by $\delta = \delta \rho_\phi/\rho_\phi = \delta_0$, and then this formalism allows us to work directly with one of the physical quantities of interest in cosmology.

\begin{figure}[h!]
    \begin{minipage}[b]{0.55\textwidth}
        \centering \includegraphics[width=\linewidth]{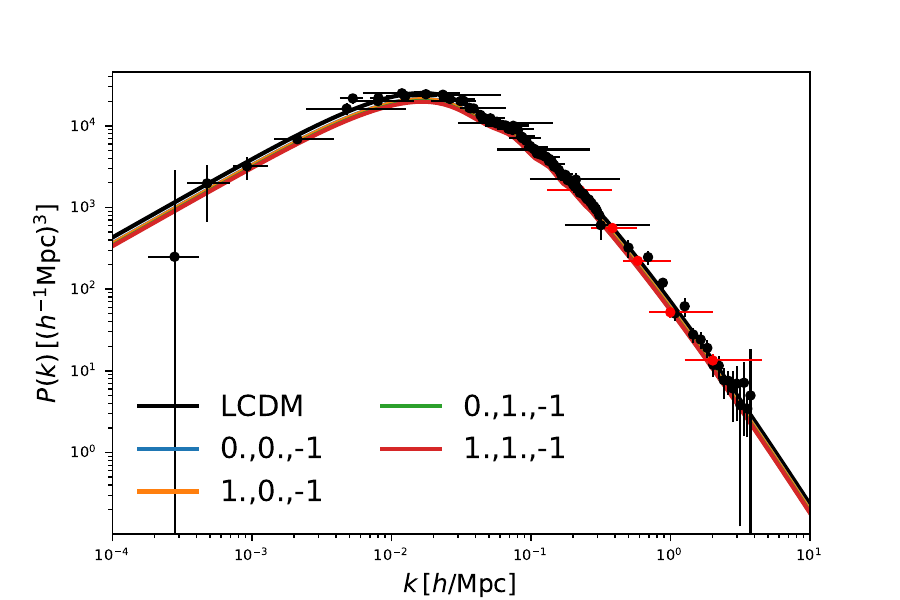}
    \end{minipage}
    \hfill
    \begin{minipage}[b]{0.55\textwidth}
        \centering \includegraphics[width=\linewidth]{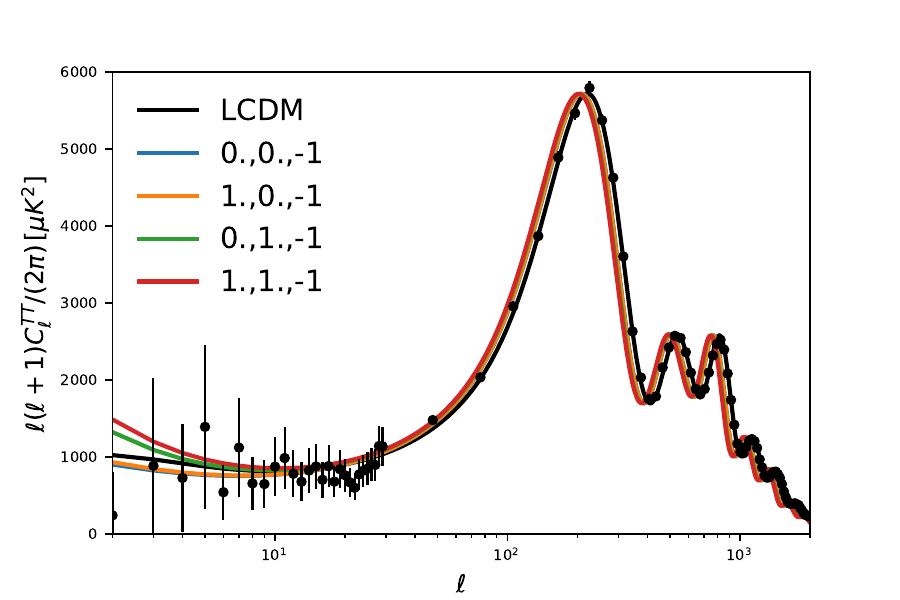}
    \end{minipage}
    \caption{The CMB anisotropies and the matter power spectrum (MPS) predicted by the tracker quintessence model, with the corresponding $\Lambda$CDM case also displayed for comparison in each panel.}\label{fig:quint2}
\end{figure}

Here, $k_J^2 = a^2 H^2 y_1,$ defines the (squared) Jeans wave number associated with the scalar field, and $k_{\rm eff}^2 = k^2 - (y_2/2y) a^2 H^2 \Omega_\phi,$ is an effective wave number that incorporates potential curvature effects through the variable $y_2$. In general, scalar field perturbations are suppressed on scales smaller than the Jeans length, this in terms of the wavenumber is $k > k_J$, while larger scales may exhibit extra growth depending on the sign of $k_{\rm eff}^2$. In particular, if $k_{\rm eff}^2 < 0$, the system can develop tachyonic instabilities, which can enhance the clustering of the scalar field. In Fig.\ref{fig:quint2} the plot of matter power spectrum (MPS) (left) and CMB anisotropies (right) for the tracker solution of the quintessence model are shown using the Boltzmann solver \texttt{CLASS}.

\subsection{The case of rapidly oscillating fields}
One example in which the polar transformation is very helpful is in the treatment of a scalar field that is able to oscillate rapidly around the minimum of its scalar potential. The typical case is a parabolic potential of the form $V= (1/2) m^2_a \phi^2$, where $m_a$ is a mass parameter of the model. Using Eqs.~\eqref{eq:polar}, and since $y_2 =0$, the polar systems of this case, for the dynamics of the background and linear density variables, are
\begin{subequations}\label{eq:polar-oscillating}
\begin{align}
\theta' &= -3\sin\theta + y_{1}, \label{eq:polar_osc}\\
y_{1}' &= \frac{3}{2} \gamma_\mathrm{tot} \, y_{1} \, , \label{eq:polar_y1_osc} \\
\Omega_{\phi}' &= 3 (\gamma_\mathrm{tot} - \gamma_\phi ) \Omega_{\phi} \, , \label{eq:polar_om_osc} \\
\delta_0' &= - \left[3\sin\theta + \frac{k^2}{k_J^2}(1-\cos\theta)\right] \delta_1 + \frac{k^2}{k_J^2}\sin\theta \, \delta_0 -\frac{h'}{2}(1-\cos\theta) \, , \label{eq:polar_d0_osc} \\
\delta_1' &= -\left(3\cos\theta + \frac{k^2}{k_J^2} \sin\theta \right) \delta_1 + \frac{k^2}{k_J^2}(1+\cos\theta) \delta_0 -\frac{h'}{2} \sin\theta \, . \label{eq:polar_d1_osc}
\end{align}
\end{subequations}
The Jeans length of density perturbations is explicitly given by $k^2_J = a^2 H^2 y_1 = 2 a^2 H m_a$, since the polar variable $y_1 = 2m_a/H$ is directly connected to the mass parameter $m_a$ of the field.

The field $\phi$ can oscillate rapidly around the minimum of the parabolic potential in $\phi =0$, and from the definition of the polar variables in Eqs.~\eqref{eq:DS_xy}, this is represented by passages of $\cos \theta$ and also by $\sin \theta$, through zero as the polar variable $\theta$ starts to grow once the dominant term in Eq.~\eqref{eq:polar_osc} is $y_1$. This happens once $y_1 \gg 1$, which means that $m_a/H \gg 1$, which leads to the standard assumption that rapid oscillations should start to occur once $m_a \gg H$.

This is also shown as oscillations of the equation of state $w_\phi$ in Eq.~\eqref{eq:polar_EoS}, and then it can be shown that on average, within a Hubble time, we can assume that $\langle w_\phi \rangle = 0$. But this is equivalent to $\langle \cos \theta \rangle = 0$ and also $\langle \sin \theta \rangle = 0$, and then Eqs.~\eqref{eq:polar-oscillating} become
\begin{subequations}\label{eq:polar-oscillating-rapid}
\begin{align}
\theta' &\simeq y_1 \, , \label{eq:polar_osc-rapid} \\
y_{1}' &= \frac{3}{2} \gamma_\mathrm{tot} \, y_{1} \, , \label{eq:polar_y1_osc-rapid} \\
\Omega_{\phi}' &\simeq 3 (\gamma_\mathrm{tot} - 1 ) \Omega_{\phi} \, , \label{eq:polar_om_osc-rapid} \\
\delta_0' &\simeq - \frac{k^2}{k_J^2} \delta_1 - \frac{h'}{2} \, , \label{eq:polar_d0_osc-rapid} \\
\delta_1' &\simeq \frac{k^2}{k_J^2} \delta_0 \, . \label{eq:polar_d1_osc-rapid}
\end{align}
\end{subequations}
Even after the start of rapid oscillations, there remains a scale-dependent growth of the density perturbations through the terms $k^2/k^2_J$ in Eqs.~\eqref{eq:polar_osc-rapid}, which in general are suppressed for small scales for which $k^2/k^2_J \gg 1$. Note that in the limit $k_J \to \infty$, equivalent to $m_a \to \infty$, such scale-dependence disappears and Eqs.~\eqref{eq:polar-oscillating-rapid} give the same results as that of a cold dark matter fluid. More details about the treatment of scalar field models with a stage of rapid oscillations under the polar system see~\cite{Urena-Lopez:2015gur,Cedeno:2017sou,Urena-Lopez:2019kud,Urena-Lopez:2019xri,LinaresCedeno:2021aqk}.

\subsection{Advantages of the alternative formalism}
Although the standard formulation with variables $(x,y)$ has proven extremely successful, the geometric interpretation of the phase space is not always transparent. In particular, the scalar field energy density and its equation of state are encoded in nonlinear combinations of the dynamical variables, which may obscure the physical meaning of trajectories. Moreover, tracking and scaling solutions, which play a central role in scalar field cosmology, do not always appear in a geometrically simple form in the $(x,y)$ plane. This can make the analysis of initial conditions and asymptotic regimes less intuitive.

As we can see from the clear classification of the critical points, the polar dynamical system approach offers several advantages over the standard $(x,y)$ formulation. First, the phase space is naturally compact, with the physically meaningful region bounded by $0 \leq \Omega_\phi \leq 1$. Second, the geometric meaning of trajectories is greatly clarified: radial motion describes the growth or decay of the scalar field energy density, while angular motion governs changes in the equation of state.

The polar formulation is also particularly well suited for studying tracking behavior and dependence on initial conditions. Since $\theta$ directly controls $w_{\phi}$, the approach allows for a clear visualization of how cosmological solutions evolve toward attractor regimes across a wide range of scalar field potentials \cite{Urena-Lopez:2020npg,LinaresCedeno:2021aqk,Roy:2023vxk}. This is also an advantage when dealing with the case of models with rapidly oscillating fields, without the need to consider an effective fluid description for the evolution of the physical variables.

\section{Hyperbolic Dynamical Systems Formulation}

Although the formulation of polar dynamical systems provides a compact and geometrically transparent description of canonical scalar field cosmologies, it may still be insufficient to capture the full global dynamics in situations where the scalar field is phantom $\epsilon = -1$ in nature. To address these limitations, a further generalization based on hyperbolic transformations has been proposed.




\begin{align}
\frac{\kappa \dot{\phi}}{\sqrt{6}H} &= \Omega_\phi^{1/2}\sinh\left(\frac{\theta}{2}\right), 
\qquad
\frac{\kappa V^{1/2}}{\sqrt{3}H} = \Omega_\phi^{1/2}\cosh\left(\frac{\theta}{2}\right), 
\end{align}

In terms of these variables, the equations in Eq.(\ref{eq:DS_xy}) for the $\epsilon =-1$ the Klein--Gordon equation can be recast as the autonomous dynamical system
\begin{subequations}\label{eq:hyperbolic}
\begin{align} 
\theta' &= -3\sinh\theta - y_1, \label{eq:hyperbolic_th}
\\
y_1' &= \frac{3}{2}(1+\gamma_{\text{tot}})y_1 
+ \frac{1}{2}\sinh\left(\frac{\theta}{2}\right)y_2, \label{eq:hyperbolic_y1}
\\
\Omega_\phi' &= 3(\gamma_{\text{tot}}-\gamma_\phi)\Omega_\phi. \label{eq:hyperbolic_om}
\end{align}
\end{subequations}
Here, a prime denotes differentiation with respect to the number of $e$-folds $N=\ln(a/a_i)$. The phantom field equation of state parameter takes the simple form
\begin{equation}
\gamma_\phi = \frac{\rho_\phi + p_\phi}{\rho_\phi} = 1 - \cosh\theta.
\end{equation}

\subsection{Critical Points and Physical Interpretation}
The dynamical system given in Eqs.~(\ref{eq:hyperbolic}) can be closed once the auxiliary variable $y_2$ is specified as a function of $y_1$ and $y$. Like the quintessence field, one can consider the general parameterization proposed in Eq.(\ref{eq:y2-polynomial}). The critical points are found by setting $\theta' = 0$, $y_1' = 0$, and $\Omega_\phi' = 0$ in the system Eq.(\ref{eq:hyperbolic}). From Eq.~(\ref{eq:hyperbolic_th}) we immediately obtain
\begin{equation}
y_{1c} = -3\sinh\theta_c.
\end{equation}
Substituting this relation into Eq.~(\ref{eq:hyperbolic_y1}), and using the parametrization~(\ref{eq:y2-polynomial}), yields the additional algebraic condition
\begin{equation}
\gamma_{\text{tot}} 
- \frac{\alpha_0}{9}\Omega_{\phi c}
+ \frac{\alpha_1}{3}\Omega_{\phi c}^{1/2}\sinh\left(\frac{\theta_c}{2}\right)
- 4\alpha_2\sinh^2\left(\frac{\theta_c}{2}\right) = 0.
\end{equation}

As in the case of quintessence, these solutions can be classified according to their physical properties.

 Scalar field dominated solution:
One possibility corresponds to $\Omega_{\phi c}=1$, for which Eq.~(10c) requires $\gamma_{\phi c}=\gamma_{\text{tot}}$. 
A particularly simple solution arises for $\theta_c=0$, implying $\gamma_{\phi c}=-1$, which describes a de Sitter-like attractor. 
Another solution is obtained for $\theta_c\to -\infty$, corresponding to $\gamma_{\phi c}\to -\infty$, and represents a kinetically dominated phantom regime.

Matter-scaling solution:
A second possibility arises from Eq.~(10c) when $\gamma_{\phi c}=\gamma_{\text{tot}}$, leading to solutions in which the phantom field mimics the background equation of state.

Field dominated Solutions: The phantom field dominated solutions are those for which $\Omega_{\phi_c}=1$.

\paragraph{Phantom Tracker Solutions:}
We now focus on the tracker solution for the phantom field. Like the quintessence field, the tracker condition for the phantom field can be obtained as 
\begin{equation}
\gamma_{\phi c} = -\frac{\gamma_{\text{tot}}}{2\alpha_2},
\end{equation}
which plays the role of the tracker condition for phantom models. For physically meaningful solutions, $\gamma_{\phi c}$ must remain finite and real, which requires $\alpha_2>0$.  A representative example is provided by power-law phantom potentials of the form $V(\phi)=M^{4-p}\phi^{p},$ for which the parameter $p$ is related to $\alpha_2$ through $p=\frac{2}{1+2\alpha_2}.$

Inverse power--law potentials correspond to $\alpha_2>0$, ensuring the existence of a tracker solution. In contrast, when $0<\alpha_2<\infty$, the phantom field evolves toward the minimum of the potential while in the tracker regime. More generally, the tracker condition can extend beyond the quadratic parameterization for some finite function $g$. Under this condition, the tracker equation reduces to
\begin{equation}
\gamma_{\text{tot}} + g\!\left(\sinh\frac{\theta_c}{2}\right) = 0.
\end{equation}

Any valid solution of this equation defines a generalized tracker behavior for the phantom field. Hence, phantom tracker solutions arise naturally for a broad class of potentials, provided the effective contribution of the potential curvature remains finite in the limit of vanishing scalar field energy density.

It is worth highlighting that, in contrast to the case of quintessence, genuine scaling solutions with $\gamma_{\phi}=\gamma_{\text{tot}}$ and $0<\Omega_\phi<1$ generally do not exist for phantom fields unless the background fluid itself behaves as a cosmological constant or possesses phantom-like properties. This difference reflects the distinct dynamical structure of phantom cosmology.

\subsection{Determination of Initial Conditions}
Similarly to polar system one can use the approximation of the thawing and subdominant dark energy at the early universe and approximate the initial conditions, see~\cite{LinaresCedeno:2021aqk} for more details. But here we discus  the estimation of the initial condition of the  hyperbolic variable under the tracker assumption. The initial values of the hyperbolic variables using the tracker  are given by
\begin{align}
\cosh\theta_i &= 1 + \frac{2}{3\alpha_2}, 
\quad 
y_{1i} = -3\sinh\theta_i, \quad \Omega_{\phi i} &= A\, a_i^{4\left(1+\frac{1}{2\alpha_2}\right)}
\left(\frac{\Omega_{m0}}{\Omega_{r0}}\right)^{1+\frac{1}{2\alpha_2}}
\Omega_{\phi 0}
\end{align}
where $\Omega_{r0}$, $\Omega_{m0}$, and $\Omega_{\phi 0}$ denote the current density parameters of radiation, matter, and phantom dark energy, respectively. The quantity $a_i$ represents the initial value of the scale factor, typically chosen deep in the radiation-dominated era (e.g. $a_i \sim 10^{-14}$). The constant $A$ serves as a normalization parameter. Fig.\ref{fig:quint1} (left) shows the evolution of the EoS of the tracker phantom scalar field and the evolution of the density parameters (right). 

\begin{figure}[h!]
    \begin{minipage}[b]{0.55\textwidth}
        \centering \includegraphics[width=\linewidth]{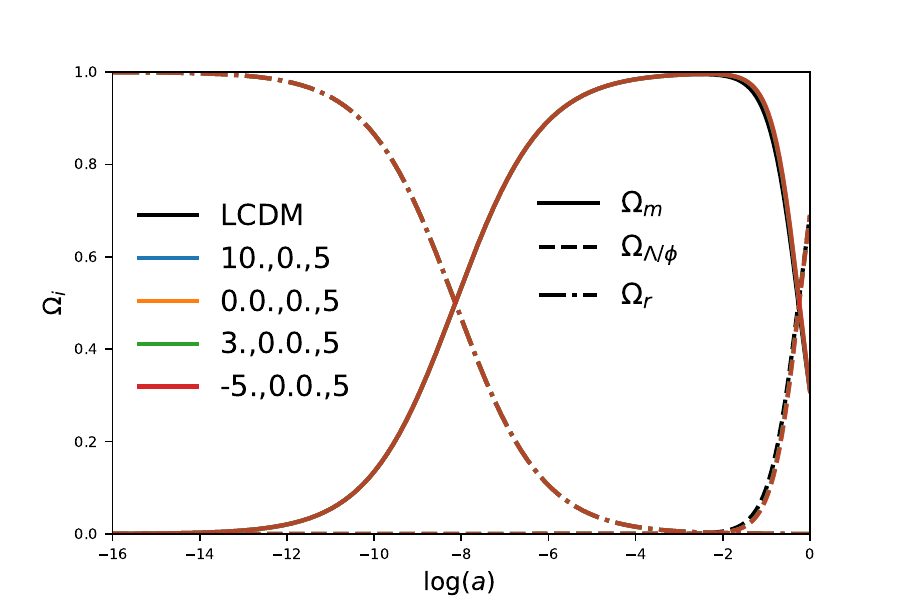}
    \end{minipage}
    \hfill
    \begin{minipage}[b]{0.55\textwidth}
        \centering \includegraphics[width=\linewidth]{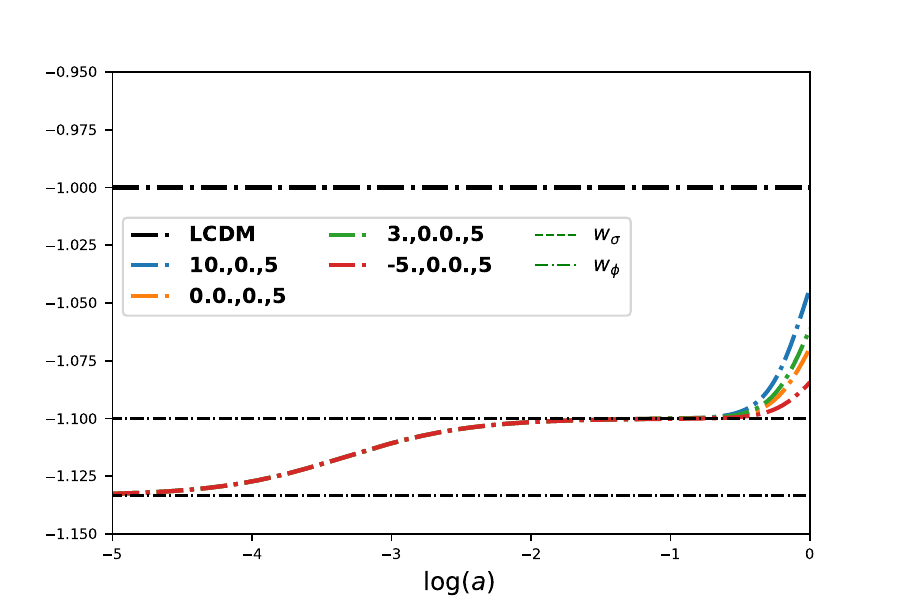}
    \end{minipage}
    \caption{(Left) The evolution of the EoS of the phantom field with the choice of the tracker initial condition for the corresponding $\alpha_i$ parameter triplets shown in the labels of the plot. The lines in blue are shown for the tracker value of the EoS during matter and radiation domination. (Right) Evolution of the density parameters of the universe for the tracker quintessence for same choice of the $\alpha_i$ parameters.}\label{fig:phantom1}
\end{figure}
\subsection{Linear Perturbations of the Phantom Field}

We now consider linear perturbations for the phantom field in the synchronous gauge, with the field decomposition $\phi(x,t) = \phi(t) + \varphi(x,t),$ where $\varphi(x,t)$ represents a small perturbation. For a Fourier mode $\varphi(k,t)$, the linearized Klein--Gordon equation for the phantom field takes the form of
\begin{equation}
\ddot{\varphi}
= -3H\dot{\varphi}
- \left( \frac{k^2}{a^2} - V_{,\phi\phi} \right)\varphi
- \frac{1}{2}\dot{h}\dot{\phi},
\end{equation}
where $k$ is the comoving wave number and $h$ is the trace of the spatial metric perturbation.

To express the perturbation equations in dynamical-system form, we introduce dimensionless variables analogous to the background hyperbolic polar parametrization,
\begin{align}
\sqrt{\frac{2}{3}} \frac{\kappa \dot{\phi}}{H} 
= -\Omega_\phi^{1/2} e^{\beta} \cosh\!\left(\frac{\vartheta}{2}\right), \quad \frac{\kappa y_1 \varphi}{\sqrt{6}} 
= -\Omega_\phi^{1/2} e^{\beta} \sinh\!\left(\frac{\vartheta}{2}\right),
\end{align}
where $\beta$ and $\vartheta$ encode the scalar field perturbations. Defining further the combinations $
\delta_0 = e^{\beta} \sinh\!\left(\tfrac{\theta}{2} + \tfrac{\vartheta}{2}\right),
\quad
\delta_1 = e^{\beta} \cosh\!\left(\tfrac{\theta}{2} + \tfrac{\vartheta}{2}\right),$
the perturbed Klein--Gordon equation can be recast as
\begin{align}
\delta'_0 &= \left[ -3 \sinh \theta - \frac{k_J^2}{k_J^2} (1 - \cosh \theta) \right] \delta_1 
- \frac{k_J^2}{k_J^2} \sinh \theta \, \delta_0 
- \frac{\hbar'}{2} (1 - \cosh \theta),
\\
\delta'_1 &= \left( -3 \cosh \theta + \frac{k_{\text{eff}}^2}{k_J^2} \sinh \theta \right) \delta_1 
- \frac{k_{\text{eff}}^2}{k_J^2} (1 + \cosh \theta) \delta_0 
+ \frac{\hbar'}{2} \sinh \theta,
\end{align}
where primes denote derivatives with respect to the number of $e$-folds $N=\ln a$. The characteristic Jeans wave number is defined very similarly as in the quintessence case as $k_J^2 = a^2 H^2 y_1,$ while the effective wave number governing the stability of perturbations is $k_{\rm eff}^2 = k^2 + (y_2/2y) a^2 H^2 \Omega_\phi .$ The variable $\delta_0$ corresponds to the fractional phantom density perturbation, $\delta_\phi = \frac{\delta\rho_\phi}{\rho_\phi} = \delta_0,$ while $k_J$ defines the associated Jeans scale. When $k_J^2>0$, perturbations are suppressed on sub-horizon scales ($k \gg k_J$). In contrast, if $k_{\rm eff}^2<0$, tachyonic instabilities can arise, potentially improving perturbation growth depending on the specific form of the potential.

\begin{figure}[h!]
    \begin{minipage}[b]{0.55\textwidth}
        \centering \includegraphics[width=\linewidth]{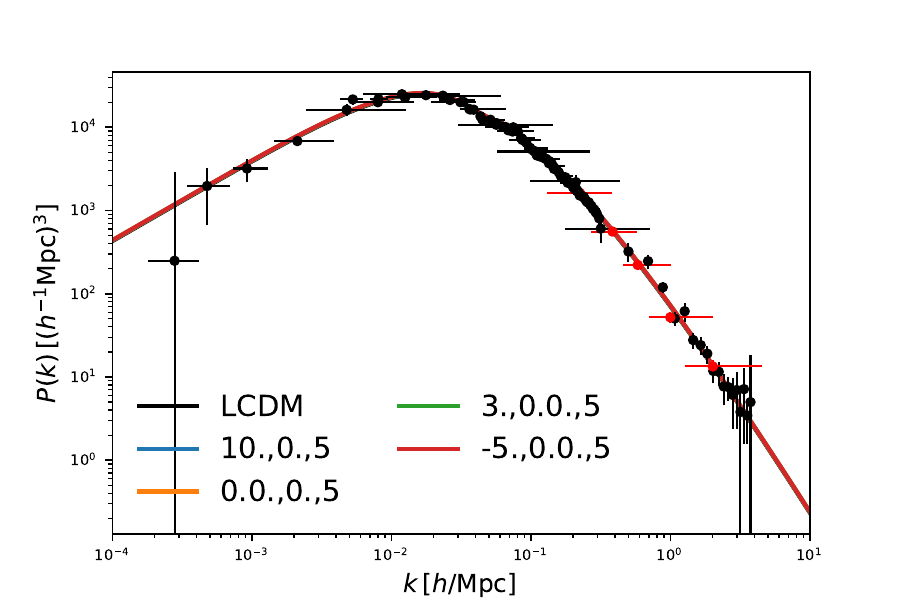}
    \end{minipage}
    \hfill
    \begin{minipage}[b]{0.55\textwidth}
        \centering \includegraphics[width=\linewidth]{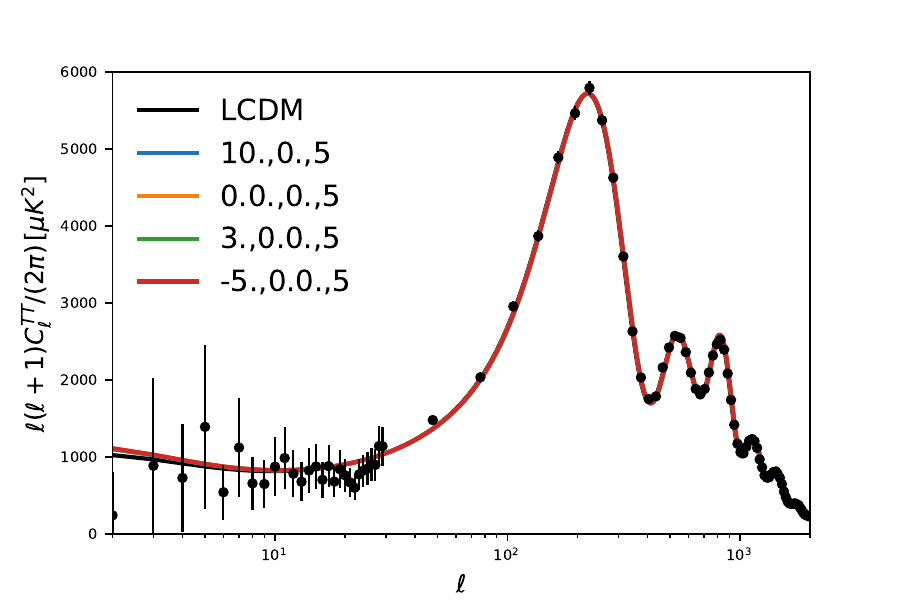}
    \end{minipage}
    \caption{The CMB anisotropies and the matter power spectrum (MPS) predicted by the tracker phantom model, with the corresponding $\Lambda$CDM case also displayed for comparison in each panel.}\label{fig:phantom2}
\end{figure}

In general, phantom density perturbations remain small as a result of the large effective sound speed of the field. Nevertheless, their inclusion is important for a consistent treatment of cosmological observables and for assessing possible instabilities associated with the phantom sector. In Fig.\ref{fig:phantom2}, the matter power spectrum (MPS) (left) and the CMB anisotropies (right) corresponding to the tracker solution of the quintessence model are presented, as computed with the Boltzmann code \texttt{CLASS}.

\section{Summary}
In this review, we have discussed the dynamical systems approach to scalar field cosmology, with emphasis on alternative formulations based on polar and hyperbolic transformations. Starting from the standard Hubble-normalized variables, we outlined how cosmological equations can be rewritten as autonomous systems whose critical points correspond to distinct cosmological regimes, such as matter domination, scalar-field domination, and scaling solutions. Linear stability analysis then determines the generic late-time behavior of the Universe.

Although the conventional formulation $(x,y)$ has been widely used, we highlighted certain limitations related to the geometric interpretation of phase space trajectories and the analysis of tracking behavior. This motivates the introduction of alternative parameterizations that provide a clearer separation between the scalar field energy density and its equation of state.

We reviewed a generalized potential parameterization that allows a broad class of scalar field models to be studied within a unified dynamical framework. In particular, the polar formulation for quintessence naturally separates density evolution and equation-of-state dynamics, leading to a compact and physically transparent phase space description. This framework facilitates the identification of tracker solutions, simplifies the estimation of initial conditions, and can be consistently extended to linear perturbations. We also discussed the hyperbolic formulation appropriate for phantom fields, which preserves the geometric clarity of the polar approach while accommodating the distinct features of phantom dynamics. 

The alternative dynamical systems formulation described above has also been extended to a variety of other dark energy models. For instance, it has been applied to interacting dark energy scenarios\cite{Roy:2018eug,Roy:2023uhc,Sahoo:2025cvz}, where energy exchange between dark matter and dark energy modifies the structure of the phase space. It has also been employed in the study of quintom models\cite{Roy:2023vxk}, in which both polar and hyperbolic formulations prove useful due to the simultaneous presence of quintessence and phantom degrees of freedom.

While the polar and hyperbolic transformations can be constructed straightforwardly for minimally coupled scalar fields with canonical or phantom kinetic terms, extending the framework to more non-trivial models can be technically challenging. For example, in theories with non-canonical kinetic terms, field-dependent couplings, modified gravity corrections, or multiple interacting scalar fields, it may not be possible to define a simple radial–angular decomposition that preserves both compactness and physical interpretability. In such cases, the construction of suitable transformed variables may become model-dependent and algebraically involved, reducing the universality of the approach.


Overall, the alternative dynamical systems formulations reviewed here provide a versatile and geometrically intuitive framework for analyzing scalar field cosmologies at both background and perturbative levels, offering complementary insights beyond the standard approach.

\bibliographystyle{unsrt}
\bibliography{reference}

\end{document}